# Dark Matter Particles with Low Mass (and FTL)

Xiaodong Huang
*Department of Mathematics, University of California, Los Angeles*
*Los Angeles, CA 90095, U.S.A.*

Wuliang Huang
*Institute of High Energy Physics, Chinese Academy of Sciences*
*P.O. Box 918(3), Beijing 100049, China*

**Abstract:** From the observed results of the space distribution of quasars and the mass scale sequence table, we deduced the existence of superstructures (feeble dark structure) with mass scale of $10^{19}$ solar mass, as well as the lightest stable fermion with mass of $10^{-1}$ eV, in the universe. From the observed results of ultra-high energy primary cosmic ray spectrum ("knee" and "ankle"), quasar irradiation spectrum ("IR bump" and "blue bump"), and Extragalactic Background Light ("IR peak" and "blue peak"), we deduced that the mass of the neutrino is about $10^{-1}$eV and the mass of the fourth stable elementary particle ($\delta$) is about $10^0$eV. While neutrino $\nu$ is related to electro-weak field, the fourth stable elementary particle $\delta$ is related to gravitation-"strong" field, and some new meta-stable baryons may appear near the *TeV* region. Therefore, a twofold standard model diagram is proposed, and involves some experiment phenomena: The new meta-stable baryons' decays produce $\delta$ particles, which are helpful in explaining the Dijet asymmetry phenomena at LHC of CERN, the different results for the Fermilab's data peak, etc; However, according to the (B-L) invariance, the sterile "neutrino" about the event excess in MiniBooNe is not the fourth neutrino but rather the $\delta$ particle; We think that the $\delta$ particles are related to the phenomenon about neutrinos FTL, and that anti-neutrinos are faster than neutrinos. FTL is also related to cosmic inflation, singular point disappearance, a finite universe, and abnormal red shift of SN Ia. Besides, the dark matter particles with low mass are helpful in explaining missing solar neutrinos, the CMB angular power spectrum measured by WMAP etc. Some experiments and observations are suggested, especially about the measurement for the speed of gravitational wave $c'$. $c'$ and c, in physics, represent the limit speeds of moving particles made by different categories of matter with different Lorentz factors. Lorentz transformation is compatible with FTL. This will be helpful to look for new particles.

Dark matter in the universe may consist of stable elementary particles with low mass (dark matter particles' mass ($m_d$) $<< m_p$).[1],[2] We use order of magnitude estimations to discuss this topic based on observations and experiments in different fields of astronomy and physics.

## The space distribution of quasars and the superstructures in the universe

In Ref [3], from the quasar number count as a function of redshift, $N(z)$, there are several peaks between $z = 0$ and $z = 5$. There are strong peaks in $N(z)$ at $z \approx 0.24$, 1.2, and 1.8. Using the results of the Fourier spectral analysis in that paper, we roughly adopted the redshift values of the peaks and made $d(z)$ calculations as follows:

| $i$ | 1 | 2 | 3 | 4 | 5 | 6 |
|---|---|---|---|---|---|---|
| $z_i$ | 0.24 | 0.5 | 0.8 | 1.2 | 1.8 | 3.0 |
| $d_i$ (Gpc) | 0.86 | 1.55 | 2.15 | 2.75 | 3.40 | 4.22 |
| $\Delta d_i$ (Gpc) | 0.86 | 0.69 | 0.60 | 0.60 | 0.65 | 0.82 |

Where $d_i$ is the distance for $z_i$, $\Delta d_i = d_i - d_{i-1}$ and $d_0 = 0$. The average value of $\Delta d_i$ is 0.7 Gpc (using the $N(z)$ data for quasi-stellar objects in Ref [4], we have an average $\Delta d_i$ of ~0.6 Gpc). This implies there are superstructures with Gpc length scale in the universe, and the related mass scale ($\sim \rho_c \cdot (\Delta d)^3$) is about $10^{19}$ solar mass.

## Large Number, Mass Scale Sequence, and Lightest Dark Matter Particle

In the mass scale sequence table of Ref [1] and [5] (see appendix I), we supposed that the maximum mass scale $M_F$ is about $A^4 m_p$ (~$10^{19}$ solar mass) followed by $m_{pl} \sim A m_p$, $m_p \sim \tilde{A}^2 m_p$, $M_{star} \sim A^3 m_p$. Then, the mass of the lightest stable fermion (non-baryonic dark matter) $m_d \sim A^{-0.5} m_p$ (~$10^{-1} eV$) can be obtained from the table, in which $A$ is the large number[6],[7],[8]: $A = \sqrt{\hbar c / G m_p^2} \sim 10^{19}$.

It is obvious that the superstructure's mass deduced from the space distribution of quasars is on the same order of magnitude as $M_F$ in the mass scale sequence table. Thus, the postulation of $M_F \sim A^4 m_p$ is correct.

## Neutrino Mass

Dark matter particles exist everywhere in the universe, and also surround the earth. The ultra-high energy primary cosmic ray spectrum (UEPCRS) measured on the surface of the earth can also reflect the interaction between primary cosmic rays and dark matter particles. As previously discussed [2],[9],[10], the "knee" of the UEPCRS can be explained by the interaction between the primary cosmic ray with energy $E_p \sim 10^{15} eV$ and the neutrinos with mass $m_\nu \sim 10^{-1} eV$ ($p + \bar{\nu} \to n + \bar{e}$). Since $m_\nu \sim m_d$, the lightest dark matter particle discussed above is the neutrino.

The host galaxies of quasars mainly are spiral galaxies, which are abundant in dark matter particles with low mass (see Appendix II). Thus, the annihilation of neutrinos and anti-neutrinos under the super-strong electromagnetic field in quasars produce IR photons ($\sim 10^{-1} eV \sim 10^{13-14} Hz$), which are related to the small IR peak in the IR bump of the quasar irradiation spectrum. [11],[12],[13],[14]

The annihilation process of $\nu$ and $\bar{\nu}$ occurs not only in quasars, but also in any place with super-strong electromagnetic field in the universe, such as galactic nuclei, neutron stars, central region of stars, black holes, etc. This means that the $10^{13-14}$ Hz IR radiation can occur at different $z$ values across the entire sky and make imprints in the Extragalactic Background Light (EBL). Thus, the small peak of EBL around $10^{-1} eV$ is also related to the annihilation of $\nu$ and $\bar{\nu}$. [15],[16]

**The fourth stable elementary particle $\delta$**

Since neutrino mass is $\sim 10^{-1} eV$, we have $\Omega_\nu \sim 10^{-2}$. Since the blue bump of the quasar irradiation spectrum and the blue light peak of EBL are around $10^0 eV$, we postulate that there is another type of dark matter particle (it can be named the $\delta$ particle) with mass of about $10^0 eV$ to provide $\Omega_\delta \sim 10^{-1}$. It is well known that dark matter particles must be stable particles. There are three types of stable particles in the universe: electron $e$ (electromagnetic interaction), neutrino $\nu$ (weak interaction), and proton $p$ (strong interaction). All of these particles are fermions. It is rational to guess that $\delta$ is the fourth stable elementary particle, which is also a fermion and connected to gravitational interaction. Then, the "ankle" of the UEPCRS can be explained by the interaction between primary cosmic ray with energy $E_p \sim 10^{17-18} eV$ and the $\delta$ particle with mass $m_\delta \sim 10^0 eV$ ($p + \delta/\bar{\delta} \to n/\bar{n} + p$). There may appear some signs of $p + \bar{p} \to n/\bar{n} + \bar{\delta}/\delta$ at LHC of CERN in the future. [1]

Since there are super-strong gravitational fields in quasars, $\delta$ particles and $\bar{\delta}$ particles annihilate in these super-strong gravitational fields and produce gravitons: $\delta + \bar{\delta} \to$ graviton + graviton. The gravitons heat the surrounding medium to produce blue light,[17] which is related to the blue bump of the quasar irradiation spectrum. The annihilation process of $\delta$ and $\bar{\delta}$ is not restricted to quasars, but rather can occur in any place with super-strong gravitational field in the universe. This forms the cosmic blue light peak in EBL. [11],[12],[13],[14]

In summary, from the "ankle" and "knee" of UEPCRS, the "blue bump" and "IR bump" of quasar irradiation spectrum, and the "blue peak" and "IR peak" of EBL, one can deduce that there are $\delta$ particles with mass of $\sim 10^0$eV and $\nu$ particles with mass of $\sim 10^{-1}$eV. There exist other clues for the existence of the dark matter particles with low mass, such as the missing solar neutrinos, the flatness of spiral galaxy rotation curves, etc. (see appendix III). The dark matter particles with low mass involve some things beyond SM and heavy WIMP, which will be discussed below.

**Beyond SM**

(1) From the table in Appendix I, it is obvious that the proton radius is connected with a "strong" gravitation constant $\widetilde{G}$. This means that $G$ "becomes" $\widetilde{G}$ in microcosm scale, and there are close relationships between gravitation interaction and proton as that between electromagnetic interaction and electron. While there are stable particle $e$ ($m_e \approx 0.5$ MeV), meta-stable particle $\mu$ ($m_\mu \approx 106$ MeV) and $\tau$ ($m_\tau \approx 1.8$ GeV) in electro-weak field, there are new meta-stable baryons with mass near TeV in gravitation-"strong" field: perhaps $p'$ (ccs) and $p''$ (ttb), which are related to $\delta'$ and $\delta''$ as $p$ (uud) related to $\delta$. All of these can be summarized as a twofold standard model diagram in Appendix IV.

(2) There are two ways to obtain $\delta$ particles (included $\delta'$ and $\delta''$) at LHC of CERN. The first way is $p + \bar{p} \to n/\bar{n} + \bar{\delta}/\delta$, but the energy threshold is too high (corresponding to "ankle"), so the event is very rare. Another way is the decay of $p'$ or $p''$ ($p' \to p\bar{\delta}\delta'$ …), which are produced in $p\bar{p}$, p-ion, ion-ion collisions. The decays produce $\delta$ particles, which are helpful in explaining the Dijet asymmetry phenomenon[18] at LHC of CERN, especially if the jets are abundant with protons.

(3) According to the Twofold Standard Model Diagram, the Fermilab's data peak[19] may relate to meta-stable particles $p'$ (or $\pi'$). The different results from D0 team[20] and CDF team at Fermilab may relate to the different effects of the new particle $\delta$ (or graviton $G$) in different detectors.

(4) The "Event Excess in the MiniBooNE Search for $\bar{\nu}_\mu \to \bar{\nu}_e$ Oscillations" (Ref [21]) supports the existence of sterile "neutrino". Since $\bar{\nu}_\mu$ has B-L=1, it is possible that the sterile "neutrino" is not the fourth (and the fifth) neutrino, but the first (and the second) $\delta$ particle. Thus, the B-L number is still invariable. This hints that the two parts of the twofold standard model are related. To check this possibility, we need to do SBL experiments above ground.

(5) As for the "neutrinos faster than light (FTL)[22]", it means that the gravitational signature propagation is faster than light and the $\delta$ particle propagation is also faster than light. Because of neutrino oscillation, some time the neutrinos become $\delta$ particles during their propagation and appear an average speed faster than light. If $\nu_\mu$ particles are faster (or not faster[22]) than light (underground), we think that $\bar{\nu}_\mu$ may propagate faster than $\nu_\mu$ (underground), this can be checked by experiments soon. There are more discussions about FTL in Appendix V.

(6) The twofold standard model diagram suggests a new interaction - relax interaction, which is mentioned above as "strong" interaction. The gravitational interaction would at first be unified with the relax interaction as electromagnetic interaction is at first unified with weak interaction.

**Lorentz Transformation and FTL**

From Appendix VI and IV, there are two categories of particles with different limit speeds of moving, c and $c'$, and the different Lorentz factors respectively. In the local interaction between different particles (belong to different categories), the conservation of energy and momentum is still correct. Lorentz transformation is compatible with FTL. This will be helpful to look for new particles.

## Some Experiments and Observations

(1) First of all is to measure the speed of gravitational wave above ground.

(2) Superstructures are feeble structures[1] (dark structure). The $\nu$ particles related to superstructures, and $(\delta + \nu)$ two components distributions related to the large scale structure with filament-like network. In order to check the existence of superstructures in the universe, we need to measure CMB angular power spectrum around $10^0 - 10^1$ degrees across the sky, as well as the space distribution of quasars, in more details.

(3) When accretion disks are consumed in quasars, it is possible that some point sources with "mono-color" gravitational wave ($10^{14-15} Hz$) / infrared emission ($10^{13-14} Hz$) can be observed across the sky. This will help us explore dark matter particles' mass and the energy source mechanism in quasars[23].

(4) The "knee" of the UEPCRS ($p + \bar{\nu} \to n + \bar{e}$, $n \to p\bar{\nu}e$) suggests an excess of cosmic ray electrons and positrons at energies $\gtrsim 10^0 TeV$ [24]. If there are $\delta$ particles, an excess at energies $\sim 10^{2-3} TeV$ is related to the "ankle" of
UEPCRS: $p + \delta/\bar{\delta} \to n/\bar{n} + p$, $n/\bar{n} \to p\bar{\nu}e / \bar{p}\nu e$. All will be observed in the future.

## Appendix I: Mass scale sequence table

| radius scale | mass scale | Classical Black Hole Radius (CBHR) | Free Stream Scale (FSS) |
|---|---|---|---|
| Planck radius $r_{pl}$ | Planck mass $m_{pl} = A\, m_p$ | "Planck" CBHR $r_{pl} \approx G\, m_{pl}/c^2$ | |
| Proton radius $r_p = A\, r_{pl}$ | Proton mass $m_p = \widetilde{A}^2 m_p$ | "strong" CBHR $r_p \approx \widetilde{G} m_p/c^2$ | |
| compact star radius $r_{star} = A^2 r_{pl}$ | star mass $M_{star} = A^3 m_p$ | star CBHR $r_{star} \approx G\, M_{star}/c^2$ | proton FSS $M_{star} \approx m_{pl}^3/m_p^2$ |
| compact superstructure radius $r_F = A^3 r_{pl}$ | superstructure mass $M_F = A^4 m_p$ | superstructure CBHR $r_F \approx G\, M_F/c^2$ | dark matter FSS $M_F \approx m_{pl}^3/m_d^2$ |

*In the table, $r_{pl} = \sqrt{G\hbar/c^3}$, $m_{pl} = \sqrt{\hbar c/G}$, $A = \sqrt{\hbar c/G m_p^2} = m_{pl}/m_p = M_F/M_{star} \approx 10^{19}$, $\widetilde{A}^2 = A^2/(A^2) = 1$, and $\widetilde{G} = (A^2) G$. We will see $M_{cr} = A^5 m_p$ and $M_u = A^6 m_p$ in Appendix V, which extend mass scale sequence.*

## Appendix II: Dark matter in quasars

In Ref [25], we discussed galaxies with two constituents: dark matter and baryonic matter ($d + B$, $m_d \ll m_B$). In that paper, we introduced a parameter "n" which can be used to distinguish spiral (S) galaxy and elliptical (E) galaxy: n > 0 for S galaxies; n < 0 for E galaxies. From calculation, it is obvious that a galaxy contains more dark matter if n > 0. Therefore, S galaxies contain more dark matter than E galaxies. Comparing number count functions $N(z)$ of S and E galaxies[26] with $N(z)$ of quasars,[3] one can see that the $N(z)$ function of S galaxies is more similar to that of quasars. Thus, one of the quasar's characteristics is that it contains abundant dark matter (and anti-dark matter) with low mass.

On the other hand, the fact that both S and E galaxies' $N(z)$ have several peaks[26] means that both types of galaxies do not have a homogeneous space distribution in a superstructure. S and E galaxies mainly occupy outer region and central region respectively. In a superstructure, the light constituent $d$ is enriched in outer regions and the heavy constituent $B$ is enriched in the central region. This means that S galaxies and quasars are mainly distributed in the outer region of a superstructure.

## Appendix III: Some Signs for Dark Matter Particles with Low Mass

The mechanism of the annihilation of dark matter particles and anti-dark matter particles is helpful in explaining missing solar neutrinos: a portion of the solar neutrinos $\nu_e$ is annihilated with dark matter anti-neutrinos $\bar{\nu}_e$ at the central region of the sun and "missed"[23]. Also, the annihilation of $\delta$ and $\bar{\delta}$ may introduce a new type of energy resource[23].

We discussed the dark matter particles with low mass in Ref [25]. The massive neutrinos were used to explain the flatness of spiral galaxy rotation curves. From the calculation of 21 samples of S galaxies, the calculated values of neutrino mass are in the range of 5eV – 33eV. According to the analysis in this paper, the dark matter particles that cause the "flatness" are not the neutrinos, but rather the $\delta$ particles.

We think that the CMB angular power spectrum measured by WMAP etc could be related to two main density waves[2] and their secondary waves. The two main density waves are produced in dark matter $\nu$ and dark matter $\delta$ respectively. For finite universe (Ref [1],[2]), the peaks of this power spectrum may reflect the mass spectrum of $\nu$ and $\delta$ particles.

The $N(z)$ of SN Ia[27],[28],[29] also reflects the existence of superstructures, but the peak's period is about two times of that for quasars.

## Appendix IV: Twofold Standard Model Diagram

We suggest the twofold Standard Model Diagram as follows:

| | | | | | | | |
|---|---|---|---|---|---|---|---|
| $u$ | $c$ | $t$ | $\gamma$ | $u'$ | $c'$ | $t'$ | $G$ |
| $d$ | $s$ | $b$ | $g$ | $d'$ | $s'$ | $b'$ | $g'$ |
| $\nu_e$ | $\nu_\mu$ | $\nu_\tau$ | $Z^0$ | $\delta$ | $\delta'$ | $\delta''$ | $Z'$ |
| $e$ | $\mu$ | $\tau$ | $W^\pm$ | $p$ | $p'$ | $p''$ | $W'$ |

Where $G$ is graviton; Z', W' may be the gauge bosons about a new type of interaction (relax interaction), which is related to $\delta$ particles as weak interaction is related to $\nu$ particles. $u'$ and $d'$ are related to a stable particle ($u'u'd'$) with mass of TeV or a stable particle with mass of ~MeV (electron). The speed of $\gamma$ is c, the speed of $G$ is c'; c'>c. The "desert" of the particle physics is not barren.

## Appendix V: FTL

The Speed of Light $c$ is a local space-time physical quantity, so:

(1) We assume there is a critical cosmic density $\rho_{cr}$ in the early universe. When cosmic density $\rho > \rho_{cr}$ during the evolution of the universe, the speed of light $c$ becomes FTL ($\bar{c}$). If $\bar{c}$ increases together with $\rho$ and $\rho/\bar{c}^2 = \rho_{cr}/c^2 =$ Constant, the evolution equation of the universe has a simple form: $\dot{R}^2 \propto R^2$ ($\dot{R} = \frac{dR}{d\tau}$, $\tau = \xi \cdot t$ and $\xi = \frac{\bar{c}(t)}{c}$). That means the universe is in an epoch of inflation.

(2) Using the afore-mentioned hypothesis to the collapse process of black hole, we get a minimum collapse radius $r_{min}$: $r_{min}^2 \propto \frac{1}{G}(\frac{c^2}{\rho_{cr}})$ for black hole, and the black hole does not become a singular point. It hints that our universe may come from a minimum $R_u$ with a finite mass $M_u$ (for example, at $\rho = \rho_{cr}$ point, $R_{cr} \sim r_{star}$ and $M_{cr} \sim A^2 M_{star} \sim A^5 m_p$; then $R_u \sim r_p$ and $M_u \sim A^6 m_p$).

(3) According to the discussion in Ref [30], $(\hbar c), G, A$ are constants. Since $(\hbar c)$ is a constant, if $c \to \sim \infty$, then $\hbar \to \sim 0$ and $r_{pl} \to \sim 0$, but $m_{pl} = const$. Therefore, we think there would be a world with dilute "u particles" before Big Bang. The u particle has mass $\sim M_u$ and radius $\sim R_u$. However, there are individual superstructures in our universe (such as CMB cold spot, Great Attractor…), which may be induced from the interaction between "u particles" and the effect of finite universe (Ref [1],[2]).

(4) There are some local weak FTL regions in observable universe, where the abnormal red shift appeared, for example, SN Ia region. It means that our universe does not seem to undergo an accelerating expansion.

# Appendix VI: A Table for Mass Tree

We extend mass scale sequence to mass tree (Ref [2] and Ref [1]). Now the mass tree has a result as follows:

| n | $m_n$ | $M_n$ | $R_n$ |
|---|---|---|---|
| 0 | $m_{pl}$ | $m_{pl}$ | |
| 1 | $m_1$ | $M_{LBH}$ | |
| 2 | $m_2 \sim m_p$ | $M_{star} \sim A^3 m_p$ | |
|   | $m' \sim m_e$ | $M' \sim M_{AGN}$ | |
|   | $m'' \sim m_\delta$ | $M'' \sim M_{galaxy-cluster}$ | |
| 3 | $m_3 \sim m_\nu$ | $M_{superstructure}$ | |
| 4 | $m_4$ | $M_4$ | $R_4 \sim A \cdot r_p$ |
| 5 | $m_5$ | $M_5 \sim A^3 M_{star}$ | $R_5 \sim r_p$ |
| 6 | $m_6$ | $M_6$ | $R_6 \sim r_p$ |
| 7 | $m_7$ | $M_7$ | $R_7 \sim r_p$ |
| 8 | $m_8 \sim m_A$ | $M_8 \sim A^3 \cdot A^3 M_{star}$ | $R_8 \sim r_p$ |

In this table there are stable particles $p, e, \delta, \nu$, where $p$ and $e$ are "bright matter"; $\delta$ and $\nu$ are dark matter. They can be expressed in Twofold / Dual SM model (arXiv: 0804.2680v6 / arXiv: 1003.5208v4) and belonged to two difference categories of matter.

---

Email address:  huangwl39@yahoo.com, huangwl@ihep.ac.cn, xhuang@ucla.edu